\documentclass[a4paper,12pt,twoside]{article}
\usepackage{amssymb,amsmath,amsthm}
\usepackage[ps,dvips,curve,knot,frame]{xy}

\renewcommand{\k}{\Bbbk}
\newcommand{\N}{\mathbb{N}}

\newcommand{\R}{\mathbb{R}}
\newcommand{\C}{\mathbb{C}}
\DeclareMathOperator{\id}{id}
\newcommand{\bcat}{\mathcal{B}}
\newcommand{\Xs}{{X^*}}
\newcommand{\hatXs}{\widehat{X^*}}
\newcommand{\tens}{\otimes}
\newcommand{\qint}[1]{[#1]_q}
\newcommand{\bint}[1]{[#1]_{\psi}}
\newcommand{\binta}[1]{[#1]'_{\psi}}
\newcommand{\bintb}[1]{[#1]''_{\psi}}
\newcommand{\defeq}{:=}
\DeclareMathOperator{\End}{\mathrm{End}}
\DeclareMathOperator{\cop}{\Delta}
\DeclareMathOperator{\cou}{\epsilon}
\DeclareMathOperator{\antip}{\mathrm{S}}
\DeclareMathOperator{\ev}{\mathrm{ev}}
\DeclareMathOperator{\evx}{\widehat{\mathrm{ev}}}
\DeclareMathOperator{\coev}{\mathrm{coev}}
\DeclareMathOperator{\diff}{\mathrm{diff}}
\DeclareMathOperator{\bZ}{\mathcal{Z}}
\DeclareMathOperator{\bZr}{\widetilde{\mathcal{Z}}}
\DeclareMathOperator{\Order}{\mathcal{O}}
\newcommand{\act}{\triangleright}
\newcommand{\Uhsl}[1]{{U_h(\mathfrak{sl}_{#1})}}
\newcommand{\Uqsl}[1]{{U_q(\mathfrak{sl}_{#1})}}
\newcommand{\SLq}[1]{{SL_q(#1)}}
\newcommand{\SUq}[1]{{SU_q(#1)}}
\newcommand{\Sq}[1]{{S_q^{#1}}}
\newcommand{\CZ}{{U(1)}}
\newcommand{\Ru}{\mathcal{R}}
\newcommand{\Al}{{{}^H\!A}}

\newcommand{\Alr}{{{}^H\!A^H}}
\newcommand{\Lc}{\mathsf{L}}
\newcommand{\Sint}{S_{\mathrm{int}}}
\newcommand{\dl}{\delta_{\mathrm{loop}}}

\renewcommand{\i}[1]{{}_{\scriptscriptstyle(#1)}}

\theoremstyle{plain}
\newtheorem{prop}{Proposition}[section]
\newtheorem{lem}[prop]{Lemma}
\newtheorem{thm}[prop]{Theorem}
\newtheorem{cor}[prop]{Corollary}

\allowdisplaybreaks

\newcommand{\sxy}[1]{{\begin{xy} #1 \end{xy}}}
\everyxy={0;<0.6mm,0mm>:<0mm,0.75mm>::0;0,}

\begin{document}
\title{\textbf{Braided Quantum Field Theory}}
\author{Robert Oeckl\footnote{email: r.oeckl@damtp.cam.ac.uk}\\ \\
Centre de Physique Th\'eorique,\\
CNRS Luminy, 13288 Marseille, France\\
\medskip
and\\
\medskip
Department of Applied Mathematics and Mathematical Physics,\\
University of Cambridge,
Cambridge CB3 0WA, UK}
\date{DAMTP-1999-82\\
29 June 1999\\
10 December 1999 (v2)\\
7 November 2000 (v3)}

\maketitle

\vspace{\stretch{1}}

\begin{abstract}
We develop a general framework for quantum field theory on
noncommutative spaces, i.e., spaces
with quantum group symmetry. We use the path integral
approach to obtain expressions for
$n$-point functions. Perturbation theory
leads us to generalised Feynman diagrams which are braided,
i.e., they have non-trivial over- and under-crossings.
We demonstrate the power of our approach by applying it to
$\phi^4$-theory on the quantum 2-sphere. We find that the basic
divergent diagram of the theory is regularised.
\end{abstract}

\vspace{\stretch{1}}

\clearpage
\section{Introduction}
The idea that space-time might not be accurately described by
ordinary geometry was expressed already a long time ago.
It was then motivated by the problems encountered
in dealing with the
divergences of quantum field theories.
An early suggestion was that spatial coordinates might in fact be
noncommuting observables \cite{Sny:quantized}.
For a long time development has been hampered by the lack of
proper mathematical tools.
Only with the advent of noncommutative geometry \cite{Con:ncgeo} and
quantum groups
have such ideas taken a more concrete form.
Quantum groups emerged in fact from the theory of integrable
models in physics and were connected from the beginning to the idea of
noncommutative symmetries in physical systems
\cite{Dri:qgroups, Jim:qgroups, Wor:cmp}. It was then also
suggested that 
they might play a role in physics at very short distances
\cite{Maj:hopfplanck}. The idea that quantum symmetry or
noncommutativity might serve as a regulator for quantum field theories
was emphasised in \cite{Maj:qreg} and \cite{Kem:ncreg}.
The persistent inability to unite quantum field theory with gravity is
a main motivation behind such considerations. In this context it is
interesting to note that noncommutative geometric structures are
emerging also in string theory \cite{CoDoSc:nctori}.
Despite progress in describing various physical
models on noncommutative spaces
(see e.g.\ \cite{Mad:ncdiffgeo, GrKlPr:4dncqft, CHMS:ftncg,
ChDePr:qftncplane}),
an approach general enough to be independent of a particular choice of
noncommutative space has been lacking.
We aim at taking a step in this direction by providing a
framework for doing quantum field theory on any noncommutative space
with quantum group symmetry.

The basic underlying idea of our approach is to take ordinary quantum
field theory, formulate it in a purely algebraic language and then
generalise in this formulation to noncommutative spaces.
It turns out that this generalisation is completely natural.
It involves no arbitrary additional input and no further choices
(except for trivial choices like taking left or right actions).
We start with two fundamental ingredients of quantum field theory,
namely the space of fields together with the group of
symmetries acting on it.
Generalising to the noncommutative context, this means that we have a
vector space of fields coacted upon by a quantum group (which we
take to mean coquasitriangular Hopf algebra) of symmetries.
Thus, the space of fields becomes an object in the
category of representations (comodules) of the quantum group,
which is \emph{braided}\footnote{Recall that a braiding means that for
two representations $V$,$W$ the intertwiner of the tensor products
$V\tens W\to W\tens V$ becomes nontrivial, i.e. different from the
flip map.}. 
I.e., we are naturally in the context of braided geometry 
\cite[Chapter~10]{Maj:qgroups}.
We emphasise that the braiding is forced on us by the requirement of
covariance under the quantum group symmetry and not introduced by
hand. 
It also turns out (at least for our example in Section~\ref{sec:phi4})
that the braiding rather than the noncommutativity
itself is crucial to achieve regularisation of a conventional theory.
This seems to have been missed out in previous works.
For previous indications that noncommutativity is not necessarily
sufficient for regularisation see e.g.\ \cite{Fil:qspace}.

We follow the path integral approach, going from
Gaussian path integrals via perturbation theory to Feynman diagrams.
In the noncommutative setting this procedure naturally leads us to
generalised Feynman diagrams that are braid diagrams, i.e., they have
nontrivial over- and under-crossings.

For an algebraically rigorous treatment we require the quantum
group of symmetries to be cosemisimple corresponding to compactness in
the commutative case. However, when aiming to regularise UV-divergences
this is not necessarily a disadvantage, since they should not be affected
by the global properties of a space.

We start out in Section~\ref{sec:formal} 
by defining normalised Gaussian integrals on braided
spaces based on \cite{KeMa:qintegration}
naturally generalising Gaussian integration on commutative spaces.
This provides us with the free $n$-point functions of a
braided quantum field theory.
Developing perturbation theory in analogy to ordinary
quantum field theory we obtain the braided analogues of Feynman
diagrams. It turns out that symmetry factors of ordinary Feynman
diagrams are resolved into different (and not necessarily equivalent)
diagrams in the braided case.

In Section~\ref{sec:homog} we consider the case where the space of
fields is a quantum homogeneous space under the symmetry quantum
group. Inspired by the conventional commutative case this
gives us a more compact description of $n$-point
functions. Furthermore, it allows for simplifications in braided
Feynman diagrams.

While our approach is somewhat formal up to this point,
Section~\ref{sec:compact} introduces a context
that allows us to work algebraically rigorously in infinite
dimensions. We need a further assumption to do this, which corresponds
in the commutative case to the space-time being compact.

Finally, in Section~\ref{sec:phi4} we deliver on the promise to
perform $q$-regularisation within
braided quantum field theory. To this end we consider $\phi^4$-theory
on the standard quantum 2-sphere \cite{Pod:qspheres}. We make use of
all the machinery developed
up to this point to show that the only basic divergence of
$\phi^4$-theory in two dimensions, the tadpole diagram, becomes finite
at $q>1$. We identify the divergence in $q$-space and suggest that it
would not depend on the conventional degree of divergence of a
diagram.

By a quantum group we generally mean a Hopf algebra
equipped with a coquasitriangular structure (see e.g.\
\cite{Maj:qgroups}.
We denote the coaction by $\cop$, the counit by
$\cou$, and the antipode by $\antip$.
We use Sweedler's notation \cite{Swe:hopfalg} $\cop a=a\i1\tens
a\i2$, etc., with summation implied. We apply the same notation to
Hopf algebras in braided categories. The braiding is denoted by
$\psi$.

While working over a general field $\k$ in
Sections~\ref{sec:formal}--\ref{sec:compact} we
specialise to the
complex numbers in Section~\ref{sec:phi4}.

\section{Formal Braided Quantum Field Theory}
\label{sec:formal}

We start out in this section by developing normalised Gaussian integration
on braided spaces leading to a braided generalisation of Wick's Theorem.
The less algebraically minded reader may find it convenient to
proceed with Section~\ref{sec:bpath} where
braided path integrals are discussed in quantum field theoretic
language, and accept the main result of Section~\ref{sec:integration}
(Theorem~\ref{thm:bWick} and its corollary) as given.

\subsection{Braided Gaussian Integration}
\label{sec:integration}

Braided categories arise as the categories of modules or comodules
over quantum groups (Hopf algebras) with quasitriangular respectively
coquasitriangular structure (see e.g.\ \cite{Maj:qgroups}).
The latter case will be the one of
interest to us later. We consider rigid braided categories,
where we have for every object $X$ a dual object $\Xs$ and morphisms
$\ev:X\tens \Xs\to\k$ (evaluation) and $\coev:\k\to \Xs\tens X$
(coevaluation) that compose to the identity in the obvious
ways. Although rigidity usually implies finite dimensionality, we shall see
later (Section~\ref{sec:compact}) how we can deal with infinite
dimensional objects.
The differentiation and Gaussian integration on braided spaces 
that we require were developed
by Majid \cite{Maj:braidedcalc} and Kempf and Majid
\cite{KeMa:qintegration} in an
$R$-matrix setting. 
(The special case of $\R_q^n$ was treated earlier in
\cite{Fio:soqharmosc}.)
We need a more abstract and basis free
formulation of the formalism so that we redevelop the notions here.
Furthermore, our Theorem~\ref{thm:bWick} goes beyond
\cite[Theorem~5.1]{KeMa:qintegration}.

Recall that a braiding on a category of vector spaces
is an assignment to any pair of vector spaces $V,W$ of
an invertible morphism $\psi_{V,W}:V\tens W\to W\tens V$.
These morphisms are required to be compatible with the tensor product
such that
$\psi_{U,V\tens W}=(\psi_{U,W}\tens\id)\circ(\id\tens\psi_{V,W})$ and
$\psi_{U\tens V,W}=(\id\tens\psi_{U,W})\circ(\psi_{U,V}\tens\id)$.
If the category is a category of modules or comodules of a quantum group
the morphisms are the intertwiners. The braiding then generalises the
trivial exchange map $\psi_{V,W}(v\tens w)=w\tens v$ which is an intertwiner
for representations of ordinary groups. In the following we simply write
$\psi$ for the braiding if no confusion can arise as to the spaces on
which it is defined.


Suppose we have some rigid braided category $\bcat$ and a vector space
$X\in\bcat$.
Essentially, we want
to define the (normalised) integral of functions $\alpha$ in the
``coordinate ring'' on $X$ multiplied by a 
Gaussian weight function $w$, i.e., we want to define
\begin{equation}
 \bZ(\alpha)\defeq\frac{\int \alpha w}{\int w} .
\label{eq:defint}
\end{equation}
First, we need to specify this
``coordinate ring''.
We identify the dual space $\Xs\in\bcat$ as the space of
``coordinate functions'' on $X$.
This corresponds to the situation in $\R^n$ where a coordinate function
is just a linear map from $\R^n$ into the real numbers.
The polynomial functions on $X$ are naturally elements of
the free unital tensor algebra over $\Xs$,
\[
 \hatXs\defeq\bigoplus_{n=0}^{\infty} \Xs^n,\quad\text{with}\quad
 \Xs^0\defeq\mathbf{1}\quad\text{and}\quad
 \Xs^n\defeq\underbrace{\Xs\tens\cdots\tens \Xs}_{n\ \text{times}},
\]
where
$\mathbf{1}$ is the one-dimensional space generated by the
identity. $\mathbf{1}$ plays the role of the constant functions
and the tensor product corresponds to the product of functions.
$\hatXs$ naturally has the structure of a braided Hopf algebra 
(a Hopf algebra in a braided category, see \cite{Maj:qgroups}) via
\[
 \cop a=a\tens 1+1\tens a,\quad\cou(a)=0,\quad\antip a=-a
\]
for $a\in\Xs$ and $\cop,\cou,\antip$
extend to $\hatXs$ as braided (anti-)algebra maps.
Explicitly, the coproduct is defined inductively by the identity
\[
 \cop\circ\cdot
 =(\cdot\tens\cdot)\circ(\id\tens\psi\tens\id)\circ(\cop\tens\cop)
\]
of maps $\hatXs\tens\hatXs\to\hatXs\tens\hatXs$.
The braided Hopf algebra structure can be thought of as
encoding translations on $X$.

To make the notion of ``coordinate ring'' more
precise, one
could perhaps consider a kind of symmetrised quotient of $\hatXs$
in analogy with the observation that coordinates commute in ordinary
geometry.
There seems to be no obvious choice for such a quotient in the
general braided case. Remarkably,
however, such a choice is not necessary. In fact,
the following discussion is entirely independent of any relations,
as long as they preserve the (graded) braided Hopf algebra structure.

The next step is the introduction of differentials \cite{Maj:braidedcalc}.
The space of coordinate differentials should be dual to the space $\Xs$
of coordinate functions. We just take $X$ itself and define
differentiation on $\Xs$
by the pairing
$\ev:X\tens \Xs\to\k$ in $\bcat$. To extend differentiation to the whole
``coordinate ring'' $\hatXs$, we note that the coproduct encodes
coordinate translation. This leads to the natural
definition that
\[
 \diff\defeq(\evx\tens\id)\circ(\id\tens\cop):X\tens\hatXs\to\hatXs
\]
is differentiation on $\hatXs$. Here,
$\evx$ is the trivial extension of $\ev$ to $X\tens\hatXs\to\k$,
i.e., 
$\evx|_{X\tens \Xs^n}=0$ for $n\neq 1$. We also use the more intuitive
notation $\partial(a)\defeq\diff(\partial\tens a)$ for $\partial\in X$ and
$a\in\hatXs$.
Let $\partial\in X$ and $\alpha,\beta\in\hatXs$. The definition of
$\evx$ gives at once
\[\evx(\partial\tens\alpha\beta)=\evx(\partial\tens\alpha)\cou(\beta)
 +\evx(\partial\tens\beta)\cou(\alpha) .\]
Using that the coproduct is a braided algebra map, we obtain the
\emph{braided Leibniz rule}
\begin{equation}
 \partial(\alpha\beta)=\partial(\alpha)\beta
 +\psi^{-1}(\partial\tens\alpha)(\beta) .
\label{eq:Leibniz}
\end{equation}
Iteration yields
\[\partial(\alpha)=(\ev\tens\id^{n-1})
 (\partial\tens\bint{n}\,\alpha) ,\]
where $n$ is the degree of $\alpha$ and
\[\bint{n}\defeq\id^n+\psi\tens\id^{n-2}+\cdots +\psi_{n-2,1}\tens\id
 +\psi_{n-1,1}\]
is a \emph{braided integer}. We adopt the convention of writing
$\psi_{n,m}$ for the braiding between $\Xs^n$ and $\Xs^m$ (respectively
$\psi^{-1}_{n,m}$ for the inverse braiding).

As in \cite{KeMa:qintegration}
we view the Gaussian weight $w$ formally as an element of $\hatXs$ and
define its differentiation via an isomorphism
\begin{equation}
 \gamma:X\to \Xs\qquad\text{so that}\qquad
 \partial(w)=-\gamma(\partial) w\qquad
 \text{for}\quad \partial\in X .
\label{eq:gamma}
\end{equation}
This expresses the familiar notion that differentiating a Gaussian
weight yields a coordinate function times the Gaussian weight.
$\gamma$ should accordingly be thought of as defining a braided
analogue of the quadratic form in the exponential of the
weight.

Also familiar from ordinary Gaussian integration is the fact
that integrals of total differentials vanish. That is, we require
\begin{equation}
\int \partial(\alpha w)=0\qquad\text{for}\qquad
 \partial\in X, \alpha\in\hatXs .
\label{eq:totint}
\end{equation}
It turns out that the three rules (\ref{eq:Leibniz}),
(\ref{eq:gamma}), and
(\ref{eq:totint}) completely determine the integral (\ref{eq:defint}).

Remarkably, the statement that the Gaussian integral of a polynomial
function can be expressed solely in terms of Gaussian integrals of
quadratic
functions still holds true in the braided case. This generalises what
is known in quantum field theory as Wick's Theorem.
To state it, we need another set of 
braided integers $\binta{n}:\Xs^n\to\Xs^n$ with
\begin{equation}
 \binta{n}\defeq \id^n+\id^{n-2}\tens\psi^{-1}+\cdots+\psi^{-1}_{1,n-1},
\end{equation}
which are related to the original ones by
$\binta{n}=\psi_{1,n-1}^{-1}\circ\bint{n}$.
We also require the corresponding \emph{braided double factorials}
$\binta{2n-1}!!:\Xs^{2n}\to\Xs^{2n}$ with
\begin{equation}
 \binta{2n-1}!!\defeq (\binta{1}\tens\id^{2n-1})\circ
 (\binta{3}\tens\id^{2n-3}) \circ\cdots\circ (\binta{2n-1}\tens\id) .
\end{equation}

\begin{thm}[Braided Wick Theorem]
\label{thm:bWick}
\begin{gather*}
 \bZ|_{\Xs^2}=\ev\circ\psi\circ(\id\tens\gamma^{-1}) ,\\
 \bZ|_{\Xs^{2n}}=(\bZ|_{\Xs^2})^n\circ \binta{2n-1}!! ,\quad
 \bZ|_{\Xs^{2n-1}}=0 ,\quad\forall n\in\N .
\end{gather*}
\end{thm}
\begin{proof}
For $\alpha\in\hatXs$ and $a\in\Xs$ we have
\[\alpha a w=-\alpha\diff(\gamma^{-1}(a)\tens w)=
 -\diff(\psi(\alpha\tens\gamma^{-1}(a))w)
 +(\diff\circ\psi(\alpha\tens\gamma^{-1}(a)))w
\]
using the differential property (\ref{eq:gamma}) of $w$ and the
braided Leibniz rule (\ref{eq:Leibniz}).
Applying $\bZ$, we can ignore the total differential and obtain
\begin{equation}
  \bZ(\alpha a)=\bZ(\diff\circ\psi(\alpha\tens\gamma^{-1}(a))) .
\label{eq:Za}
\end{equation}
This gives us immediately
\[\bZ(a)=0\qquad\text{and}\qquad
 \bZ(a b)=\ev\circ\psi(a\tens \gamma^{-1}(b))
\]
for $b\in\Xs$. We rewrite (\ref{eq:Za}) to find
\[
\begin{split}
\bZ|_{\Xs^n} &=\bZ|_{\Xs^{n-2}}\circ\diff\circ(\gamma^{-1}\tens\id^{n-1})
 \circ\psi_{n-1,1}\\
&=\bZ|_{\Xs^{n-2}}\circ(\ev\tens\id^{n-2})\circ(\gamma^{-1}\tens\bint{n-1})
 \circ\psi_{n-1,1}\\
&=(\ev\tens\bZ|_{\Xs^{n-2}})\circ(\gamma^{-1}\tens\bint{n-1})
 \circ\psi_{n-1,1}\\
&=(\ev\tens\bZ|_{\Xs^{n-2}})\circ\psi_{n-1,1}
 \circ(\bint{n-1}\tens\gamma^{-1})\\
&=(\bZ|_{\Xs^2}\tens\bZ|_{\Xs^{n-2}})\circ(\id\tens\psi_{n-2,1})
 \circ(\bint{n-1}\tens\id)\\
&=(\bZ|_{\Xs^{n-2}}\tens\bZ|_{\Xs^2})\circ\psi_{2,n-2}^{-1}
 \circ(\id\tens\psi_{n-2,1})
 \circ(\bint{n-1}\tens\id)\\
&=(\bZ|_{\Xs^{n-2}}\tens\bZ|_{\Xs^2})\circ(\psi_{1,n-2}^{-1}\tens\id)
 \circ(\bint{n-1}\tens\id)\\
&=(\bZ|_{\Xs^{n-2}}\tens\bZ|_{\Xs^2})
 \circ(\binta{n-1}\tens\id) ,
\end{split}
\]
which gives us a recursive definition of $\bZ$ leading to the formulas
stated.
\end{proof}

Another set of the braided integers
\begin{align*}
 &\bintb{n}
 \defeq\id^n+\psi^{-1}\tens\id^{n-2}+\cdots+\psi_{1,n-1}^{-1}\\
\text{with}\quad
 &\bintb{2n-1}!! \defeq (\id\tens\bintb{2n-1})\cdots
(\id^{2n-3}\tens\bintb{3})(\id^{2n-1}\tens\bintb{1})
\end{align*}
serves to formulate the dual version of the theorem.

\begin{cor}
\label{cor:int}
Let $\bZ^k\in X^k$ denote the dual of $\bZ|_{\Xs^k}$. Then
\begin{gather*}
 \bZ^{2}=\psi\circ(\gamma^{-1}\tens\id)\circ\coev,\\
 \bZ^{2n}=\bintb{2n-1}!!\, (\bZ^2)^n,\quad
 \bZ^{2n-1}=0,\quad\forall n\in\N,
\end{gather*}
\end{cor}
\begin{proof}
This is obtained from Theorem~\ref{thm:bWick} by reversing of
arrows or equivalently by turning 
diagrams upside down in the diagrammatic language of braided
categories.
\end{proof}

\subsection{Braided Path Integrals}
\label{sec:bpath}

The $n$-point function of an ordinary quantum field theory with
action $S$, evaluated at
$(x_1,\ldots,x_n)$ is given by the path integral\footnote{The
Euclidean signature of the action is chosen for definiteness and
does not imply a restriction to Euclidean field theory.}
\[
 \langle \phi(x_1)\cdots\phi(x_n)\rangle
 = \frac{\int \mathcal{D}\phi\, \phi(x_1)\cdots\phi(x_n) e^{-S(\phi)}}
        {\int \mathcal{D}\phi\, e^{-S(\phi)}} .
\]
This is really the normalised integral of the functional
$\phi\mapsto \phi(x_1)\cdots\phi(x_n)$ with weight
$w(\phi)=e^{-S(\phi)}$ over the space $X$ of classical fields
of the theory. The parameters $x_i$ denote here
points in space-time as well as additional internal field indices.

For the non-interacting
theory the action $S$ is replaced by the free action $S_0$.
The path integral is then a Gaussian
integral and the decomposition of $n$-point functions into
2-point functions (propagators) is governed by Wick's theorem.
Generalising to braided spaces (when the symmetry group
is allowed to be a quantum group) we are
in the framework of
Section \ref{sec:integration}. Then, the value of an $n$-point function
is still given in terms of values of 2-point functions (propagators).
This is the result of Theorem~\ref{thm:bWick} which generalises
Wick's Theorem.
The (unevaluated) $n$-point function $\bZ^n$ itself is an element in the
$n$-fold tensor product $X^n$ of the space of fields $X$ and we write
\[
 \bZ^n(x_1,\dots,x_n)=\langle \phi(x_1)\cdots\phi(x_n)\rangle_0 ,
\]
the index $0$ indicating that we deal with the free theory.
The decomposition of $\bZ^n$ into propagators $\bZ^2$ is given by
Corollary~\ref{cor:int}, which is Theorem~\ref{thm:bWick} in dual form,
i.e., for ``unevaluated'' functions.

The connection between the map $\gamma$ determining the (unevaluated)
propagator according to Theorem~\ref{thm:bWick} (Corollary~\ref{cor:int})
and the free action
in ordinary quantum field theory is as follows.
Let $\partial$ be some differential with respect to the space of
fields. The definition of $\gamma$ in (\ref{eq:gamma})
corresponds to
\[
 (\partial(e^{-S_0}))(\phi)=-(\gamma(\partial))(\phi)
 e^{-S_0(\phi)} ,
\]
in ordinary quantum field theory.
Thus we obtain
\begin{equation}
 (\gamma(\partial))(\phi)=(\partial S_0)(\phi) .
\label{eq:gamact}
\end{equation}

To determine interacting $n$-point functions, we use the
same perturbative techniques as in ordinary quantum field theory.
For $S=S_0+\lambda \Sint$ with coupling constant
$\lambda$, we expand
\begin{align*}
 \bZ^{n}_{\mathrm{int}}(x_1,\dots,x_n) & =
 \langle \phi(x_1)\cdots\phi(x_n)\rangle \\
 & = \frac{\int \mathcal{D}\phi\, \phi(x_1)\cdots\phi(x_n) 
 (1-\lambda \Sint(\phi)+\dots) e^{-S_0(\phi)}}
        {\int \mathcal{D}\phi\, (1-\lambda \Sint(\phi)+\dots)
  e^{-S_0(\phi)}} \\
 & = \frac{\langle \phi(x_1)\cdots\phi(x_n)\rangle_0
    -\lambda \langle\phi(x_1)\cdots\phi(x_n)\Sint(\phi)\rangle_0
    +\dots}
   {1-\lambda \langle\Sint(\phi)\rangle_0
    +\dots} .
\end{align*}
For $\Sint$ of degree $k$ we can write
\[
 \langle \phi(x_1)\dots\phi(x_n)\Sint(\phi)\rangle_0
 = ((\id^n\tens\Sint)\bZ^{n+k})(x_1,\dots,x_n)
\]
etc.\ by viewing $\Sint$ as a map $X^k\to\k$.
Then, removing the explicit evaluations we obtain
\begin{equation}
\bZ^{n}_{\mathrm{int}}=
 \frac{\bZ^n-\lambda (\id^n\tens\Sint)(\bZ^{n+k})
 +\frac{1}{2}\lambda^2 (\id^n\tens\Sint\tens\Sint)(\bZ^{n+2k})+\ldots}
 {1-\lambda \Sint(\bZ^{k})
 +\frac{1}{2}\lambda^2 (\Sint\tens\Sint)(\bZ^{2k})+\ldots},
\label{eq:bnpoint}
\end{equation}
an expression for the interacting $n$-point function valid
in the general braided case.
Vacuum contributions cancel as usual.
Note that we have used the ordinary exponential expansion for the
interaction and not,
say, a certain braided version.
The latter might be more natural if, e.g., one wants to look at identities
between diagrams of different order.
However, we shall not consider this issue here.

\subsection{Braided Feynman Diagrams}
\label{sec:bFeyn}

\begin{figure}
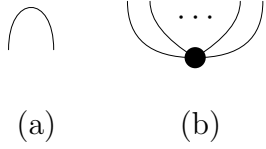

\begin{center}
\begin{tabular}{ccc}
\input{fig_arch}
& &
\input{fig_vertex}
\\ \\
(a) & & (b)
\end{tabular}
\caption{Propagator (a) and vertex (b).}
\label{fig:arch_vertex}
\end{center}
\end{figure}

\begin{figure}
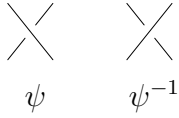

\begin{center}
\begin{tabular}{ccc}
\input{fig_over}
& &
\input{fig_under}
\\
$\psi$ & & $\psi^{-1}$
\end{tabular}
\caption{The braiding and its inverse.}
\label{fig:braiding}
\end{center}
\end{figure}

We are now ready to generalise Feynman Diagrams to our braided
setting. To do this we use and modify the diagrammatic language of
braided categories appropriately:
\begin{itemize}
\item
An $n$-point function is an element in
$X\tens\cdots\tens X$ ($n$-fold). Thus, its diagram is closed to the
top and ends in $n$ strands on the bottom. Any strand represents an
element of $X$, i.e., a field.
\item
The propagator $\bZ^2\in X\tens X$ is represented by an
arch, see Figure~\ref{fig:arch_vertex}.a.
\item
An $n$-leg vertex is a map $X\tens\cdots\tens X\to\k$. It is
represented by $n$ strands joining in a dot, see
Figure~\ref{fig:arch_vertex}.b. Notice that
the order of incoming strands is relevant.
\item
Over- and under-crossings correspond to the braiding and its inverse,
see Figure~\ref{fig:braiding}.
\item
Any Feynman diagram is built out of propagators,
(possibly different kinds of) vertices, and strands with crossings,
connecting the propagators and vertices, or ending at the bottom.
\end{itemize}
Otherwise the usual rules of braided diagrammatics apply.
Notice that in contrast to ordinary Feynman diagrams all external legs
end on one line (the bottom line of the diagram)
and are ordered. This is necessary due to the possible non-trivial braid
statistics in our setting.
For the case of trivial braiding we can relax this and shift
the external legs around as well as change the order of strands at
vertices so as to obtain ordinary Feynman diagrams in
more familiar form.

\begin{figure}
\begin{center}
\input{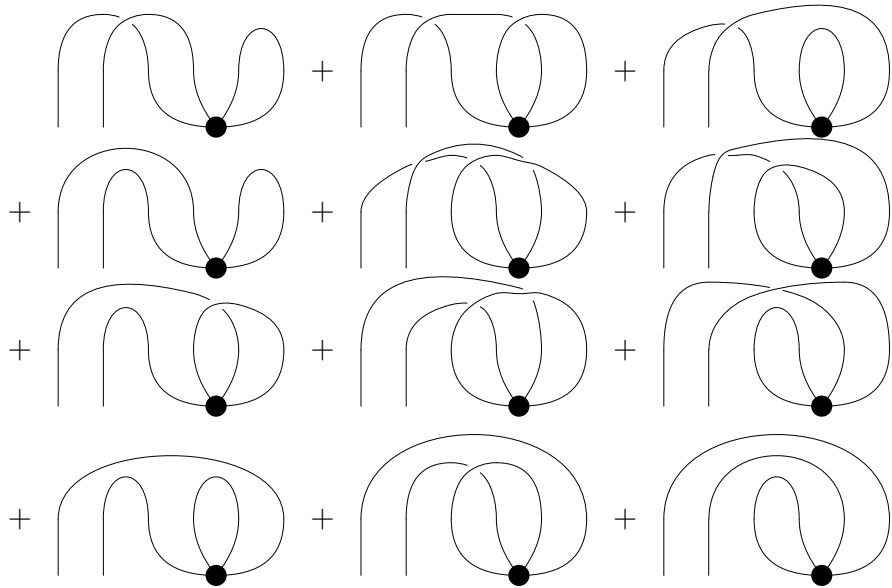}
\caption{Interacting 2-point function of $\phi^4$-theory at order 1.}
\label{fig:int2}
\end{center}
\end{figure}

\begin{figure}
\begin{center}
\input{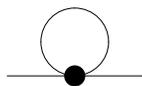}
\caption{Tadpole diagram of ordinary $\phi^4$-theory.}
\label{fig:tadpole}
\end{center}
\end{figure}

The diagrams for the free $2n$-point functions can be read off
directly from Corollary~\ref{cor:int}. The crossings are encoded in
the braided integers $\bintb{j}$.
Figure~\ref{fig:free4} shows
for example the free $4$-point function and Figure~\ref{fig:free6} the
free $6$-point function.
For the interacting $n$-point functions we use formula
($\ref{eq:bnpoint}$) to obtain the diagrams. $\Sint$ gives us the
vertices. Consider for example the $2$-point function in Euclidean
$\phi^4$-theory. 
To order $\lambda$ we get
\begin{equation}
 \bZ^{2}_{\mathrm{int}}=
  \bZ^2-\lambda \left((\id^2\tens\Sint)(\bZ^6)
  -\bZ^2\tens\Sint(\bZ^4)\right)
  +\Order(\lambda^2) .
\label{eq:phi4}
\end{equation}
$\Sint$ is just the map
$\phi_1\tens \phi_2\tens \phi_3\tens\phi_4\mapsto
\int \phi_1\phi_2\phi_3\phi_4$. To obtain the diagrams at order
$\lambda$ we start by drawing the free 6-point function
(Figure~\ref{fig:free6}) and attach to
the 4 rightmost strands of each diagram a 4-leg vertex
(Figure~\ref{fig:arch_vertex}.b). Those diagrams are 
generated by the first term in brackets of (\ref{eq:phi4}).
We realise
that the first three of our diagrams are vacuum diagrams which are
exactly 
cancelled by the second term in the brackets. The remaining 12
diagrams are shown in Figure~\ref{fig:int2}. In ordinary quantum field
theory they all correspond to the same diagram: The tadpole diagram,
see Figure~\ref{fig:tadpole}. However, not all of them are
necessarily different, as we shall see in Section~\ref{sec:diatech}.

\begin{figure}
\begin{center}
\input{fig_free4}
\caption{Free 4-point function.}
\label{fig:free4}
\end{center}
\end{figure}

\begin{figure}
\begin{center}
\input{fig_free6}
\caption{Free 6-point function.}
\label{fig:free6}
\end{center}
\end{figure}

\section{Braided QFT on Homogeneous Spaces}
\label{sec:homog}

In ordinary quantum field theory fixing one point of an $n$-point
function still allows to recover the whole $n$-point function. 
Thus, we can reduce an $n$-point function to a function of just $n-1$
variables. This
is simply due to the fact that any $n$-point function is invariant
under the isometry group $G$ of the space-time $M$ and $G$
acts transitively on $M$. In this case $M$ is a homogeneous space
under $G$ and we can 
make the above statement more precise in the following way.

\begin{lem}
\label{lem:cired}
Let $G$ be a group and $K$ a subgroup of $G$.
For any $n\in\N$ there is an isomorphism of coset spaces
\[ \rho_n:
  (\underbrace{K\backslash G\times\cdots\times K\backslash G}_{n\
   \text{times}})/G \cong
  (\underbrace{K\backslash G\times\cdots\times K\backslash G}_{n-1\
   \text{times}})/K
\]
given by $\rho_n:[a_1,\ldots,a_n]\mapsto [a_1 a_n^{-1},\ldots,a_{n-1}
a_n^{-1}]$ for $a_i\in K\backslash G$. Its inverse is given by
$\rho_n^{-1}:[b_1,\ldots,b_{n-1}]\mapsto [b_1,\ldots,b_{n-1},e]$ for
$b_i\in K\backslash G$, where $e$ denotes the equivalence class of the
identity in $K\backslash G$.
If $G$ is a topological group (i.e., it is a
topological space and multiplication and inversion are continuous),
then equipping the coset spaces with the induced topologies makes 
$\rho_n$ into a homeomorphism.
\end{lem}

If space-time is an ordinary manifold we can obviously do the same
trick in braided quantum field theory. More interestingly, however, we
can extend it to noncommutative space-times.

\subsection{Quantum Homogeneous Spaces}
\label{sec:qhom}

Lemma~\ref{lem:cired}
generalises to the quantum group case. To see this we first
recall the notion of a quantum homogeneous space.

Suppose we have two Hopf algebras $A$ and $H$ together with a Hopf
algebra surjection $\pi:A\to H$. This induces
coactions $\beta_L=(\pi\tens\id)\circ\cop$ and
$\beta_R=(\id\tens\pi)\circ\cop$ of $H$
on $A$, making $A$ into a left and right $H$-comodule algebra. Define
$\Al$ to be the left $H$-invariant subalgebra of $A$, i.e.,
$\Al=\{a\in A|\beta_L(a)=1\tens a\}$. 
We have $\cop \Al\subseteq \Al\tens A$ since
$(\beta_L\tens\id)\circ\cop=(\id\tens\cop)\circ\beta_L$. This makes
$\Al$ into a right $A$-comodule (and $H$-comodule) algebra. 
Observe also that $\pi(a)=\cou(a) 1$ for $a\in\Al$.
$\Al$ is
called a right
\emph{quantum homogeneous space}. Define the left quantum homogeneous
space $A^H$ correspondingly.
Due to the anti-coalgebra property of the antipode we find $\antip
\Al\subseteq A^H$ and $\antip A^H\subseteq \Al$. If the
antipode is invertible, the inclusions become equalities.

\begin{prop}
\label{prop:ired}
In the above setting with invertible antipode the map
\[\rho_n:
(\underbrace{\Al\tens\cdots\tens\Al}_{n\ \text{times}})^A 
\to (\underbrace{\Al\tens\cdots\tens\Al}_{n-1\ \text{times}})^H
\]
given by $\rho_n=(\id^{n-1}\tens\cou)$ for $n\in\N$
is an isomorphism. Its inverse is
$(\id^{n-1}\tens\antip)\circ\beta^{n-1}$,
where $\beta^{n-1}$ is the right coaction of $A$ on $\Al$ extended to
the $(n-1)$-fold tensor product.
\end{prop}
\begin{proof}
Let $a^1\tens\cdots\tens a^n$ be an element of
$(\Al\tens\cdots\tens\Al)^A$. In particular,
\[ a^1\i1\tens\cdots\tens a^n\i1\tens a^1\i2\cdots a^n\i2
 = a^1\tens\cdots\tens a^n\tens 1 .\]
Applying the antipode to the
last component and multiplying with the $n$-th component
we obtain
\begin{equation}
 a^1\i1\tens\cdots\tens a^{n-1}\i1\tens\cou(a^n)
  \antip(a^1\i2\cdots a^{n-1}\i2)
 = a^1\tens\cdots\tens a^n .
\label{eq:ireda}
\end{equation}
Thus,
$(\id^{n-1}\tens\antip)\circ\beta^{n-1}\circ(\id^{n-1}\tens\cou)$ is
the identity on $(\Al\tens\cdots\tens\Al)^A$.
On the other hand, applying 
the inverse antipode and then $\pi$ to the
last component of (\ref{eq:ireda})
we get
\[ a^1\i1\tens\cdots\tens a^{n-1}\i1\tens\cou(a^n)
  \pi(a^1\i2\cdots a^{n-1}\i2)
 = a^1\tens\cdots\tens a^{n-1}\tens \cou(a^n)1 .\]
This is to say that $a^1\tens\cdots\tens a^{n-1}\cou(a^n)$ is indeed
right $H$-invariant.

Conversely, it is clear that $(\id^{n-1}\tens\cou)\circ
(\id^{n-1}\tens\antip)\circ\beta^{n-1}
=(\id^{n-1}\tens\cou)\circ\beta^{n-1}$ is the identity.
Now take $b^1\tens\cdots\tens b^{n-1}$ in
$(\Al\tens\cdots\tens\Al)^H$.
Its image under $\beta^{n-1}$ is
\begin{equation}
 b^1\i1\tens\cdots\tens b^{n-1}\i1\tens b^1\i2\cdots b^{n-1}\i2 .
\label{eq:iredb}
\end{equation}
Applying $\pi$ to the last component we get
\[
 b^1\i1\tens\cdots\tens b^{n-1}\i1\tens \pi(b^1\i2\cdots b^{n-1}\i2)
 = b^1\tens\cdots\tens b^{n-1}\tens 1
\]
by right $H$-invariance. Applying $\beta^{n-1}\tens\id$ we arrive at
\begin{align*}
 b^1\i1\tens\cdots\tens b^{n-1}\i1\tens b^1\i2\cdots b^{n-1}\i2
  \tens\pi(b^1\i3\cdots b^{n-1}\i3)\\
 = b^1\i1\tens\cdots\tens b^{n-1}\i1
  \tens b^1\i2\cdots b^{n-1}\i2 \tens 1 .
\end{align*}
We observe that this is the same as applying $(\id^{n-1}\tens\beta_R)$
to (\ref{eq:iredb}). Thus, the last component of (\ref{eq:iredb})
lives in $A^H$ and the application
of the antipode sends it to $\Al$ as required. That the result is right
$A$-invariant is also clear by the defining property of the antipode.
\end{proof}

To make use of the result we assume our space $X$ of fields to be a
quantum homogeneous space under a quantum group
(coquasitriangular Hopf algebra) $A$ of symmetries.
(Note that coquasitriangularity implies invertibility of the
antipode.) That is, together with $A$ we have another Hopf algebra $H$
and a Hopf 
algebra surjection $A\to H$. We then assume that the algebra of fields
is the right quantum homogeneous space $X=\Al$
living in the braided category $\mathcal{M}^A$ of right $A$-comodules.

\subsection{Diagrammatic Techniques}
\label{sec:diatech}

\begin{figure}[p]
\begin{center}
\input{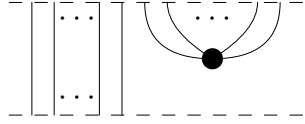}
\caption{Vertex evaluation in a diagram slice.}
\label{fig:vslice}
\end{center}
\end{figure}

\begin{figure}
\begin{center}
\input{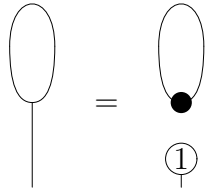}
\caption{Extracting a loop.}
\label{fig:loopex}
\end{center}
\end{figure}

\begin{figure}
\begin{center}
\input{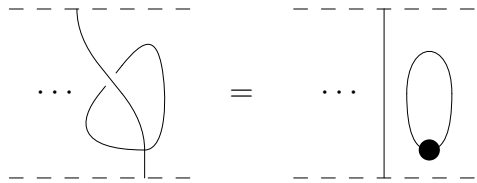}
\caption{Separating a loop in an invariant slice.}
\label{fig:loopsep}
\end{center}
\end{figure}

\begin{figure}
\begin{center}
\input{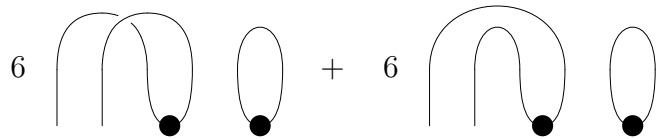}
\caption{Simplified 2-point function of $\phi^4$-theory at order 1.}
\label{fig:int2simp}
\end{center}
\end{figure}

Proposition~\ref{prop:ired}, to which we shall refer as \emph{invariant
reduction}, is not only useful to express $n$-point functions in a
more compact way, but can also be applied in the evaluation of braided
Feynman diagrams.
For this we note
that any horizontal cut of a braided Feynman diagram
lives in some tensor power of $X$ (since the only allowed strand lives
in $X$) and is invariant (since the diagram is closed at the top).
Thus, we can apply invariant reduction to it.
We shall give three examples for this, assuming vertices that are
evaluated by multiplication and subsequent integration. Here, any
quantum group invariant linear map $X\to\k$ is admissible as the
integral.
\begin{description}
\item[Vertex evaluation.]
Consider the evaluation of an $n$-leg vertex (the horizontal
slice of an 
invariant diagram depicted in Figure~\ref{fig:vslice})
with incoming elements $a_1\tens\cdots\tens a_{k+n}$.
By invariant reduction this can be expressed in two ways,
\begin{align*}
 & a_1\tens\cdots\tens a_k \int a_{k+1}\cdots a_{k+n} \\
 & = a_1\i1\tens\cdots\tens a_k\i1 \cou(a_{k+1})\cdots\cou(a_{k+n})
  \int\antip(a_1\i2\cdots a_k\i2)
\end{align*}
Depending on the circumstances each side might be easier to evaluate.
\item[Loop extraction.]
Assume that the integral on $\Al$ is normalised, $\int 1=1$.
Consider the diagram in Figure~\ref{fig:loopex} (left-hand side). It
is obviously invariant. Thus, the
single outgoing strand carries a multiple of the identity and we can
replace it by the integral followed by the identity element
(Figure~\ref{fig:loopex}, right-hand side).
\item[Loop separation.]
 We assume further that the coquasitriangular structure
 $\Ru:H\tens H\to\k$ is trivial on $\Alr$ in the sense
\begin{equation}
 \Ru(a\tens b)=\cou(a)\cou(b),\qquad\text{if}\ a\in\Alr\ \text{or}
   \ b\in\Alr .
\label{eq:Rprop}
\end{equation}
Consider now the diagram in Figure~\ref{fig:loopsep} (left-hand side)
as a horizontal slice of an
invariant diagram. According to invariant reduction we apply the counit
to the rightmost outgoing strand.
This makes the braiding trivial due
to the assumed property of $\Ru$. We can push the counit up to each
of the joining strands and disentangle them. Then proceeding as in the
previous example leads to the diagram in Figure~\ref{fig:loopsep}
(right-hand side).
Note that this works the same way for an under-crossing.
\end{description}
Let us come back to the 2-point function of $\phi^4$ theory that we
considered at the end of Section~\ref{sec:bFeyn}.
Assuming $\int 1 = 1$ and property
(\ref{eq:Rprop}) we can use loop extraction and loop separation
to simplify the order 1
diagrams of Figure~\ref{fig:int2}
considerably. The result is shown in Figure~\ref{fig:int2simp}.
Instead of 12 different diagrams we only have 2 different and much
simpler diagrams, each with a multiplicity of 6.

\section{Braided QFT on Compact Spaces}
\label{sec:compact}

\subsection{Braided Spaces of Infinite Dimension}

Up to now we have developed our approach on a
formal level insofar, that
we have not addressed the question how an
infinite dimensional space (of fields) can be treated in a braided
category.
This is certainly necessary if we want to do quantum field theory,
i.e., deal with infinitely many degrees of
freedom.
An obvious problem is the definition of the coevaluation.
It seems that we need at least a completed tensor product for this.
However, instead of introducing heavy functional analytic machinery,
we can stick with our algebraic approach given a further assumption.

Let us assume that the space of (regular) fields $X$
decomposes into a direct sum $\bigoplus_i X_i$
of countably many finite dimensional
comodules under the
symmetry quantum group $A$.
This corresponds roughly to the
classical case of the space-time manifold being compact.
In particular, it is the case if the symmetry quantum group
$A$ is cosemisimple (or classically
the Lie group of symmetries is compact, see
Section~\ref{sec:PeterWeyl} below). Denote the projection
$X\to X_i$ by $\tau_i$.

We now allow arbitrary sums of elements in $X$ given that any projection
$\tau_i$ annihilates all but finitely many summands. Similarly,
we allow infinite sums in the $n$-fold tensor product $X^n$ with the
restriction that any projection $\tau_{i_1}\tens\cdots\tens\tau_{i_n}$
yields a finite sum. To define the dual of $X$, we take the dual of
each $X_i$ and set $X^*=\bigoplus_i X_i^*$.
For each component $X_i$ we have an evaluation map $\ev_i:X_i\tens
X_i^*\to\k$ and a coevaluation map $\coev_i:\k\to X_i\tens X_i^*$ in the
usual way. We then formally define $\ev=\sum_i
\ev_i\circ(\tau_i\tens\tau_i^*)$ and $\coev=\sum_i \coev_i$.

Our definition is invariant under coactions of $A$ as it should be,
since the projections $\tau_i$ commute with the coaction of $A$.
In particular, it is invariant under braidings.

\subsection{Cosemisimplicity and Peter-Weyl Decomposition}
\label{sec:PeterWeyl}

We describe a context in which all comodules over a Hopf algebra
decompose into finite dimensional (and even simple) pieces. The
discussion here uses results of \cite{Swe:hopfalg} but is more in the
spirit of \cite[II.9]{CaSeMa:liegroups}. Assume $\k$ to be algebraically
closed, e.g., $\k=\C$.

Let $C$ be a coalgebra, $V$ a simple right $C$-comodule (i.e.\ $V$ has
no proper subcomodules) with coaction $\beta:V\to V\tens C$.
In particular, $V$ is finite dimensional. The
dual space $V^*$ is canonically a (simple) left $C$-comodule.
Denote a basis of $V$ by $\{e_i\}$, the dual basis of $V^*$ by $\{f^i\}$.
Identify the endomorphism algebra on $V$, $\End V\cong V\tens V^*$ via
$(e_i\tens f^j)(e_k\tens f^l)=\delta^j_k (e_i\tens f^l)$. We denote
the dual coalgebra by $(\End V)^*$ and identify $(\End V)^*\cong
V^*\tens V$ via $\cop (f^i\tens e_j)=\sum_k(f^i\tens e_k)
\tens(f^k\tens e_j)$.

Now consider the map $(\End V)^*\to C$ given by $f^i\tens e_j\mapsto
(f^i\tens\id)\circ\beta(e_j)$. It is an injective (since $V$ is
simple) coalgebra map. We extend this to the direct sum of all
inequivalent simple comodules. The resulting map
\[
 \bigoplus_V (\End V)^*\to C
\]
is a coalgebra injection. It is an isomorphism of coalgebras if and
only if all $C$-comodules are semisimple (i.e.\ they are direct sums
of simple ones) or equivalently if $C$ is semisimple (i.e.\ it is a
direct sum of simple coalgebras).

Assume now that $A$ is a \emph{cosemisimple} Hopf algebra, i.e., $A$
is semisimple as a coalgebra.
We write the above decomposition as
\begin{equation}
 A\cong \bigoplus_V (V^*\tens V) .
\label{eq:PeterWeyl}
\end{equation}
It is also referred to as the Peter-Weyl
decomposition, in analogy to 
the corresponding decomposition of the algebra of regular functions on
a compact Lie group.
There is a unique normalised left- and right-invariant integral 
(Haar measure) on $A$, 
given by the induced projection to the unit element in
$A$.
Note also that the antipode is invertible.

Consider a second Hopf algebra $H$ with a Hopf algebra surjection
$\pi:A\to H$. This induces a coaction of $H$ on each $A$-comodule.
For the right quantum homogeneous space we have
\begin{equation}
 \Al\cong \bigoplus_V ({}^H\!(V^*)\tens V)
\label{eq:PWhom}
\end{equation}
as right $H$-comodules.

\section{$\phi^4$-Theory on the Quantum 2-Sphere}
\label{sec:phi4}

In accordance with the motivation of braided quantum field theory as a
way of regularising ordinary quantum field theory, we replace Lie
groups of symmetries by corresponding parametric deformations.
In order to have a well defined theory in
the sense of Section~\ref{sec:compact} we make use of the Peter-Weyl
decomposition and thus restrict to compact Lie groups.
A natural choice are the standard $q$-deformations of Lie groups
with compact $*$-structure.
We specialise to $\k=\C$, although the discussion of the free
action in Section~\ref{sec:bpath} was in the spirit of real-valued
scalar field theory. This is necessary since the standard 
$q$-deformations viewed as deformations of complexifications of
compact Lie groups
do not restrict to real subalgebras for
$q\neq 1$.
However,
viewing $q$-deformation purely as a mathematical tool 
we can always restrict to $\R$ when considering physical
quantities living at $q=1$.

In the following we consider perturbative $\phi^4$-theory on the
quantum 2-sphere with $\SUq2$-symmetry
as an example of a quantum field theory on a braided space.
Ordinary $\phi^4$-theory in 2 dimensions is super-renormalisable and
has just one basic divergence: The tadpole diagram
(Figure~\ref{fig:tadpole}).
(See e.g.\ \cite{Zin:qftcrit} for a treatment of ordinary
$\phi^4$-theory.) 
We demonstrate that this diagram becomes finite for $q>1$.
Our Hopf algebra of symmetries is $\SUq2$ under which 
$\Sq2$ is a homogeneous space as a right comodule.
(We adopt the convention to denote the Hopf algebra of regular
functions by the name of the (quantum) group or space.)

\subsection{The Decomposition of $\SUq2$ and $\Sq2$}
\label{sec:suq2}

To prepare the ground we need to recall the construction of $\Sq2$ as
a quantum  homogeneous space under $\SUq2$ and the Peter-Weyl
decomposition of the latter \cite{Koo:suq2,MMNNU:suq2}. This
will enable us to apply the
machinery of the previous sections.

Recall that $\SUq2$ is the compact real form of $\SLq2$ for $q$ real
which we assume in the following.
(See Appendix~\ref{app:suq2} for the defining relations.)
It is cosemisimple and
there is one simple (right) comodule $V_l$ for each
integer dimension, conventionally labelled by a
half-integer $l$ such that the dimension is $2l+1$. 
Thus, the Peter-Weyl decomposition (\ref{eq:PeterWeyl}) is
\[
 \SUq2\cong \bigoplus_{l\in\frac{1}{2}\N_0} (V_l^*\tens V_l) .
\]

There is a Hopf $*$-algebra surjection $\pi:\SUq2\to\CZ$ 
corresponding to the diagonal inclusion in the commutative case.
(See Appendix~\ref{app:suq2} for an explicit definition of $\pi$.)
This defines the
quantum 2-sphere $\Sq2$ as the right quantum homogeneous $*$-space
${{}^{\CZ}\SUq2}$.
Under
the coaction of $\CZ$ induced by $\pi$ the comodules $V_l$ decompose into
inequivalent 
one-dimensional comodules classified by integers. (This is the usual
representation theory of $\CZ$.)
This determines up to normalisation a basis $\{v^{(l)}_n\}$ for $V_l$
with half-integers $n$ taking values $-l,-l+1,\ldots,l$.
In particular, we find
that $V_l^{\CZ}$ is one-dimensional if $l$ is integer and
zero-dimensional otherwise.
Thus, (\ref{eq:PWhom}) simplifies to
\[
 \Sq2\cong \bigoplus_{l\in\N_0} V_l
\]
as right $\SUq2$-comodules.
We write the induced (normalisation independent) basis vectors of
$\SUq2$ as 
$t^{(l)}_{i\,j}= (f^{(l)}_i\tens\id)\circ\beta(e^{(l)}_j)$ where
$f^{(l)}_n$ is dual to $e^{(l)}_n$ and $\beta:V_l\to V_l\tens\SUq2$ is
the coaction of $\SUq2$ on $V_l$.
As a subalgebra $\Sq2$ has the
basis $\{t^{(l)}_{0\,i}\}$. The bi-invariant subalgebra
$\Sq2^\CZ={}^\CZ\SUq2^\CZ$ has the basis $\{t^{(l)}_{0\,0}\}$.

Note that by construction
\[
 \cou\left(t^{(l)}_{m\,n}\right)=\delta_{m,n}\quad
 \text{and}\quad
 \cop t^{(l)}_{m\,n} = \sum_k t^{(l)}_{m\,k}\tens t^{(l)}_{k\,n} .
\]
The antipode and $*$-structure of $\SUq2$ in this basis are
\[
 \antip t^{(l)}_{m\,n}=(-q)^{m-n} t^{(l)}_{-n\,-m},
 \quad
 \left(t^{(l)}_{m\,n}\right)^*=\antip t^{(l)}_{n\,m}=
 (-q)^{n-m} t^{(l)}_{-m\,-n} ,
\]
as can be verified by direct calculation from the formulas in
\cite[4.2.4]{KlSc:qgroups}.
The normalised invariant integral (Haar measure) is simply
$\int t^{(l)}_{i\,j}=\delta_{l,0}$.
We also need its value on the product of two basis elements
\begin{equation}
\int t^{(l)}_{m\,n} t^{(l')}_{m'\,n'}=
\frac{(-1)^{m-n} q^{m+n}}{\qint{2l+1}}
\delta_{l,l'}\delta_{m+m',0}\delta_{n+n',0} .
\label{eq:intprod}
\end{equation}
This can be easily worked out considering the equation $\cou(a)=\int
a\i1\antip a\i2$ and using the invariance of the integral in the form
$b\i1\int a b\i2=\antip a\i1 \int a\i2 b$ and
$\antip b\i2 \int a b\i1=a\i2 \int a\i1 b$ on basis elements.
The $q$-integers for $q\in\C^*$ are defined as
\[
 \qint{n}\defeq\sum_{k=0}^{n-1} q^{n-2k-1}
 =\frac{q^n-q^{-n}}{q-q^{-1}} . 
\]
(The second expression is only defined for $q^2\neq 1$).

Denoting a dual basis of $\{t^{(l)}_{m\,n}\}$ by
$\{\tilde{t}^{(l)}_{m\,n}\}$, we observe
that $\SUq2^*$ becomes
an object in $\mathcal{M}^\SUq2$, the category of right comodules
over $\SUq2$
by equipping it with the coaction
$\tilde{t}^{(l)}_{m\,n}\mapsto \sum_k \tilde{t}^{(l)}_{m\,k}
\tens\antip^{-1} t^{(l)}_{n\,k}$.
We then have an evaluation map
$\ev:\SUq2\tens\SUq2^*\to\C$ and a coevaluation map
$\coev:\C\to\SUq2^*\tens\SUq2$ in the obvious way.

In the commutative case $q=1$, the basis
$\{t^{(l)}_{m\,n}\}$ becomes
the usual basis of regular
functions (i.e., matrix elements of representations) on $SU(2)$ (see
e.g.\ \cite[Chapter~6]{ViKl:repspecfunc1} to whose conventions we conform in
this case). The restriction to $\{t^{(l)}_{0\,n}\}$ recovers
nothing but (a version of) the spherical harmonics on $S^2$. In
particular, we notice that the zonal spherical functions can be
expressed in terms of Legendre polynomials
$t^{(l)}_{0\,0}(\phi,\theta,\psi)=P_l(\cos\theta)$, where
$\phi,\theta,\psi$ are the Euler angles on $SU(2)$ (see
\cite[Chapter~6]{ViKl:repspecfunc1}). From the orthogonality relation of the
Legendre 
polynomials, the fact that their only common value is at $P_l(1)=1$,
and considering that $\theta=0$ denotes a pole of $SU(2)$,
we find that the delta function at the identity of $SU(2)$ restricted
to $S^2$ can be represented as
\begin{equation}
 \delta_0(\phi,\theta)=\sum_l (2l+1)\, P_l(\cos\theta)
 =\sum_l (2l+1)\, t^{(l)}_{0\,0}(\phi,\theta) .
\label{eq:delta}
\end{equation}

Recall that a coquasitriangular structure $\Ru:H\tens H\to\k$
on a quantum group $H$ determines a braiding between right comodules
$V$ and $W$ via
\[
 \psi(v\tens w)=w\i1\tens v\i1 \Ru(v\i2\tens w\i2)
\]
for $v\in V$ and $w\in W$. (We use here Sweedler's coproduct notation
for the coaction.) 
For calculations we need the functionals $u$ and $v$ defined with
$\Ru$ as (see e.g.\
\cite{Maj:qgroups})
\begin{equation}
 u(a)\defeq\Ru(a\i2\tens \antip a\i1) , \quad
 v(a)\defeq\Ru(a\i1\tens \antip a\i2)
\label{eq:defcqtruv}
\end{equation}
for $a\in H$.
For $H=\SUq2$ in our basis they are
\begin{equation}
u(t^{(l)}_{m\,n}) =\delta_{m,n}\, q^{-2l(l+1)+2m},\quad
v(t^{(l)}_{m\,n}) =\delta_{m,n}\, q^{-2l(l+1)-2m} .
\label{eq:sl2uv}
\end{equation}
We also note that property (\ref{eq:Rprop}) is satisfied, i.e.,
\begin{equation}
 \Ru\left(t^{(l)}_{0\,0}\tens t^{(l)}_{i\,j}\right)
 = \delta_{i,j} =
  \Ru\left(t^{(l)}_{i\,j}\tens t^{(l)}_{0\,0}\right) .
\label{eq:sl2prop}
\end{equation}
See Appendix~\ref{app:cqtr} for a derivation of (\ref{eq:sl2uv}) and
(\ref{eq:sl2prop}).

\subsection{The Free Propagator}

In ordinary quantum field theory the free propagator is defined by
the free action. For a Euclidean massive real scalar field theory
on a manifold $M$ it takes the form
\[
 S_0(\phi)=\frac{1}{2}\int_M dx\,\phi(x) (m^2-\Delta_M)\phi(x) ,
\]
where $\Delta_M$ is the Laplace operator on $M$ and $m$ is the mass of
the field. Define $\Lc\defeq m^2-\Delta_M$.
Let $\{\phi_i\}$ be a basis of $X$ and $\{\phi_i^*\}$ a dual basis.
Denote the differential with respect to $\phi_i$ by $\partial_i$. We
have
\[
 (\partial_i S_0)(\phi)=\int_M dx\,\phi(x)\Lc \phi_i(x)
  = \sum_k \phi_k^*(\phi)\int_M dx\,\phi_k(x)\Lc\phi_i(x) .
\]
Comparing with equation (\ref{eq:gamact}) we obtain in the more abstract
notation of Section~\ref{sec:integration}
\begin{equation}
 \gamma=\left(\id\tens\int_M\right)\circ(\id\tens\cdot)
 \circ(\coev\tens\Lc) ,
\label{eq:defgamma}
\end{equation}
which we take as the defining equation for $\gamma$.
While initially well defined only at $q=1$ we extend it to the
noncommutative realm in the following.

First, note that at $q\neq 1$ we still have a well defined integral
on our ``manifold'' $M=\Sq2$, namely the induced Haar measure of $\SUq2$.
Next, we need an
analogue of the Laplace operator. 
By the duality of $\SUq2$ with the quantum enveloping
algebra $\Uqsl2$, a central element of
the latter defines an invariant operator on $\SUq2$-comodules.
A natural choice is the quantum
Casimir element which we define as
\[
 C_q=E F+\frac{(K-1) q^{-1}+(K^{-1}-1) q}{(q-q^{-1})^2} .
\]
Here $K$, $K^{-1}$, $E$, and $F$ are the generators of $\Uqsl2$ (see
Appendix~\ref{app:cqtr}).
$C_q$ differs from quantum Casimir elements
considered elsewhere (see e.g.\ \cite{MMNNU:suq2} or \cite{KlSc:qgroups}) 
only by a
$q$-multiple of the identity. 
The eigenvalue of $C_q$ on $V_l$
is $\qint{l}\qint{l+1}$ so that we get
exactly the (negative of the) usual Laplace operator for $q=1$.
Including a mass term we set
\[
 \Lc=C_q+m^2 .
\]
Thus, the eigenvalue of $\Lc$ on $V_l$ is
\[
 \Lc_l=\qint{l}\qint{l+1}+m^2 .
\]
We determine $\gamma$ according to (\ref{eq:defgamma}). Using
(\ref{eq:intprod}) we find
\[
 \gamma \left(t^{(l)}_{0\,i}\right) = \sum_{m,j}
   \tilde{t}^{(m)}_{0\,j}\int t^{(m)}_{0\,j}\,
    \Lc\left(t^{(l)}_{0\,i}\right)
  = \qint{2l+1}^{-1}\, \Lc_l\, (-q)^{-i}\, \tilde{t}^{(l)}_{0\,-i} ,
\]
Inverting we obtain
\[
 \gamma^{-1}\left(\tilde{t}^{(l)}_{0\,i}\right)
 = \qint{2l+1}\, \Lc_l^{-1}\, (-q)^{-i}\, t^{(l)}_{0\,-i} .
\]
Now we are ready to determine the free propagator according to
Corollary~\ref{cor:int}.
\begin{align*}
 \bZ^2 &=\sum_{l,k} (\id\tens\gamma^{-1})\circ
  \psi\left(\tilde{t}^{(l)}_{0\,k}\tens t^{(l)}_{0\,k}\right)\\
 &=\sum_{l,i,j,k}
  t^{(l)}_{0\,i}\tens \gamma^{-1}\left(\tilde{t}^{(l)}_{0\,j}\right)\,
  \Ru\left(\antip^{-1} t^{(l)}_{k\,j}\tens t^{(l)}_{i\,k}\right)\\
 &=\sum_{l,i,j}
  t^{(l)}_{0\,i}\tens \gamma^{-1}\left(\tilde{t}^{(l)}_{0\,j}\right)\,
  u\left(t^{(l)}_{i\,j}\right)\\
 &=\sum_{l,i} \qint{2l+1}\, \Lc_l^{-1}\, q^{-2l(l+1)}\, (-q)^{i}\,
  t^{(l)}_{0\,i}\tens t^{(l)}_{0\,-i} .
\end{align*}
Using invariant reduction (Proposition~\ref{prop:ired}) we find
\begin{equation}
 \bZr^2= \sum_{l} \qint{2l+1}\, \Lc_l^{-1}\,
  q^{-2l(l+1)}\, t^{(l)}_{0\,0}
\label{eq:free2red}
\end{equation}
to be the reduced form of the propagator as an element of
$\Sq2^\CZ$.
In the commutative case $q=1$ we can rewrite (\ref{eq:free2red}) as
\[
 \bZr^2|_{q=1} 
  =(m^2-\Delta)^{-1}\delta_0
\]
by comparison with (\ref{eq:delta}). This is the familiar expression
from ordinary quantum field theory.

\subsection{Interactions}

We proceed to evaluate the order 1 contribution of the
$\phi^4$-interaction to the 2-point
function. The corresponding diagrams are depicted in
Figure~\ref{fig:int2} (see Section~\ref{sec:bFeyn}).
Since the property (\ref{eq:Rprop})
holds in $\SUq2$ the diagrams simplify to those of
Figure~\ref{fig:int2simp}
(see Section~\ref{sec:diatech}). The disconnected loop comes out as
\begin{equation}
 \dl\defeq\,\sxy{
(40,0);(50,0)**\crv{(40,10)&(50,10)},
(45,-10)*\frm<4pt>{*};(40,0)**\crv{(40,-10)},
(50,0)**\crv{(50,-10)}}
\,
  =\sum_{l} \frac{\qint{2l+1}}{\qint{l}\qint{l+1}+m^2}\,q^{-2l(l+1)} .
\label{eq:loop}
\end{equation}
(Just apply the counit to (\ref{eq:free2red}).)
The connected diagram in the right-hand summand of
Figure~\ref{fig:int2simp} is (in reduced form)
\begin{align*}
 \sxy{
(30,0)**\crv{(0,15)&(30,15)},
(10,0);(20,0)**\crv{(10,10)&(20,10)},
0;(0,-10)**\dir{-},(10,0);(10,-10)**\dir{-},
(25,-10)*\frm<4pt>{*};(20,0)**\crv{(20,-10)},
(30,0)**\crv{(30,-10)}}
\,
 & = \left(\id\tens\cou\tens\int\right)\circ(\id^2\tens\cdot)\circ
 (\id\tens\bZ^2\tens\id)\circ\bZ^2 \\
 & = \sum_{l,m,i,j} \alpha_l\alpha_m\, t^{(l)}_{0\,i}
  \cou\left(t^{(m)}_{0\,j}\right)
  \int \antip t^{(m)}_{j\,0}\, \antip t^{(l)}_{i\,0}\\
 & = \sum_{l} \alpha_l^2\, \qint{2l+1}^{-1}\, t^{(l)}_{0\,0} ,
\end{align*}
with $\alpha_l\defeq\qint{2l+1}\, \Lc_l^{-1}\,q^{-2l(l+1)}$.
We have used $\bZ^2$ as reconstructed from its reduced form
(\ref{eq:free2red}), the property
$\int\circ\antip=\int$ of the integral, and (\ref{eq:intprod}).
The connected diagram in the left-hand summand of
Figure~\ref{fig:int2simp} is (in reduced form)
\begin{align*}
 \sxy{
(10,10)**\crv{(0,10)},
(20,10);(30,0)**\crv{(30,10)},
\vunder~{(10,10)}{(20,10)}{(10,0)}{(20,0)},
0;(0,-10)**\dir{-},(10,0);(10,-10)**\dir{-},
(25,-10)*\frm<4pt>{*};(20,0)**\crv{(20,-10)},
(30,0)**\crv{(30,-10)}}
\,
 & = \left(\id\tens\cou\tens\int\right)\circ(\id^2\tens\cdot)\circ
 (\id\tens\psi^{-1}\tens\id)\circ(\bZ^2\tens\bZ^2)\\
 & = \sum_{l,m,i,j,k,n} \alpha_l\alpha_m\, t^{(l)}_{0\,i}
   \cou\left(t^{(m)}_{0\,k}\right)
  \int \antip t^{(l)}_{n\,0}\, \antip t^{(m)}_{j\,0}\,
  \Ru^{-1}\left(t^{(m)}_{k\,j} \tens \antip t^{(l)}_{i\,n}\right)\\
 & = \sum_{l,m,i,j,n} \alpha_l\alpha_m\, t^{(l)}_{0\,i}
  \int t^{(m)}_{j\,0}\, t^{(l)}_{n\,0}\,
  \Ru\left(t^{(m)}_{0\,j} \tens t^{(l)}_{i\,n}\right)\\
 & = \sum_{l,m,i,j,k} \alpha_l\alpha_m\, t^{(l)}_{0\,i}
  \int t^{(m)}_{k\,0}\, t^{(l)}_{i\,0}\,
  \Ru\left(t^{(m)}_{0\,j} \tens \antip t^{(m)}_{j\,k}\right)\\
 & = \sum_{l,m,i,k} \alpha_l\alpha_m\, t^{(l)}_{0\,i}
  \int t^{(m)}_{k\,0}\, t^{(l)}_{i\,0}\,
  v\left(t^{(m)}_{0\,k}\right)\\
 & = \sum_{l} \alpha_l^2\, \qint{2l+1}^{-1}\,q^{-2l(l+1)}\,
  t^{(l)}_{0\,0} .
\end{align*}
We have also used the invariance of the integral in the form
$(\int a b\i2) b\i1=(\int a\i2 b) \antip a\i1$ in the third equality.
Thus, the (reduced) 2-point function up to order 1 comes out as
\begin{equation}
\begin{split}
 \bZr^2_{\mathrm{int}}
= & \sum_l \qint{2l+1}\,\Lc_l^{-1}\,q^{-2l(l+1)}\,t^{(l)}_{0\,0}\\
 & \left( 1- 6\,\lambda\,\dl\, \Lc_l^{-1} q^{-2l(l+1)} (1 + q^{-2l(l+1)})
  +\Order(\lambda^2)\right) .
\end{split}
\label{eq:int2}
\end{equation}
In the commutative case ($q=1$), we know that the order 1 contribution
(given by
the tadpole diagram in Figure~\ref{fig:tadpole}) is divergent. We can
easily see where this divergence comes from. The loop contribution
(\ref{eq:loop})
\begin{equation}
 \dl|_{q=1}=\sum_l \frac{2l+1}{l(l+1)+m^2}
\label{eq:classdiv}
\end{equation}
is infinite. However, at $q>1$ it becomes finite. We are truly
able to regularise the tadpole diagram.
Let us identify the divergence in $q$-space.
For $q>1$ we can find both an upper and a lower bound for
(\ref{eq:loop}) of the form
\[
 \mathsf{const}+\int_1^\infty dl\,\frac{2}{l} q^{-2 l^2} ,
\]
where $\mathsf{const}$ does not depend on $q$ (but may depend on
$m^2$). Setting $q=e^{2 h^2}$ with $h>0$ we find
\[
 \dl|_{q>1}=\frac{1}{h}+\Order(1) .
\]

The conventional divergence of (\ref{eq:classdiv}) is only logarithmic
in $l$. What would happen with higher divergences?
It seems natural to
assume that they would give rise to terms
like
\[
 \sum_l\, \qint{l}^n\, q^{-2l(l+1)} .
\]
But this converges in the domain $q>1$ for any $n$. We can even apply
the very same discussion of the divergence in $q$-space as above.
The nature of the divergence in $q$-space does not seem to be
affected by the degree of the ordinary (commutative) divergence at all.
This suggests that $q$-regularisation in our framework is powerful
indeed.

Reviewing our calculations of $\bZ^2$ and $\bZ^2_{\mathrm{int}}$ we
find that the crucial factor of $q^{-2l(l+1)}$ is caused by the
braiding. Thus, the braiding and not the mere noncommutativity
appears to be essential for the regularisation.

\subsection{Renormalisation}

Ordinarily, $\phi^4$-theory in dimension 2 is
super-renormalisable. The only basic divergent diagram is the tadpole
(Figure~\ref{fig:tadpole}).
Our approach yields a simple and diagrammatic way to renormalise it.
We have used above the loop separation technique of
Section~\ref{sec:diatech} (Figure~\ref{fig:loopsep}) to factorise the single
tadpole diagram(s) into $\phi^2$-vertex diagrams and the loop factor $\dl$.
For any given diagram we can perform the same operation for all tadpole
subdiagrams appearing in it. The remaining
diagram (with the loop factors removed) is finite at $q=1$, since the
commutative theory has no further divergences.

However, from a rigorous point of view
this procedure can only be performed if the diagram we start out with
is finite.
While we have
seen that the tadpole diagram alone becomes finite for $q>1$, it is
conceivable that certain diagrams that converge at $q=1$ would diverge
at $q>1$. This might be due to the introduction of factors like
$q^{2l(l+1)}$ into summations over $l$. The expression (\ref{eq:int2})
suggests, however, that
this does not happen, but rather that all $q$-factors introduced in
summations have negative exponent.
We shall assume this in the following.

Let us perform the usual mass renormalisation in our framework.
We introduce an extra perturbative mass term which generates diagrams
with $\phi^2$-vertices. These diagrams are then used to cancel
the corresponding diagrams where the $\phi^2$-vertices are the remnants
of the factorisation of tadpole subdiagrams. To effect the cancellation
the perturbative mass term must carry the same factor $\dl$ as the
factorised tadpoles. To compensate for the different combinatoric
multiplicity of quadratic and quartic vertices we need an extra
factor of 6 in front of the $\phi^2$-vertex. Since a mass term carries
an overall factor of $1/2$ in the action, the effective mass shift is
\[
 m^2 \to m^2-12\lambda\,\dl .
\]
Performing this (finite) mass renormalisation at $q>1$, only the
divergence-free diagrams without tadpoles remain as $q\to 1$ at any
given order in perturbation theory.

\section{Concluding Remarks}

We have presented a coherent framework for the treatment of quantum
field theory on braided spaces. In particular, we have developed a
quantum group covariant perturbation theory.

The example of $\phi^4$-theory on the quantum
2-sphere has shown that quantum deformations of symmetries do lead to
the regularisation of divergences in our approach.
This method is superior to regularisation methods such as using a lattice
or fuzzy spaces in that it does not resort to discrete approximations
with only finitely many degrees of freedom.
On the other hand it does not suffer from the crude breaking of symmetries
as many quantum field theoretic methods do (e.g.\ momentum cut-off,
dimensional regularisation, lattice).
However, symmetries are not preserved as such, but deformed
to quantum group symmetries.
Our results also suggest that divergences of arbitrary order could be
regularised in this way.

A next step would be the investigation of quantum field theories on
deformations of higher dimensional spaces to obtain more
physically interesting models.
We note in particular that quantum deformations of
Minkowski space are available (see \cite{CaScScWa:qmink, MaMe:qmink}
and \cite{LNR:kpoincare, MaRu:kappa}).
Further one would like to include internal (quantum group)
symmetries as well. In particular, this might open new possibilities
for the old idea of unifying internal and external symmetries.

In a different direction, one
might speculate that the braided Feynman diagrams obtained
from theories with $q$-deformed symmetries have interesting
number theoretic properties related to modular functions.
This is suggested by the observation of such properties for
the quantum rank of $q$-deformed enveloping algebras
\cite{MaSo:rankquantmod}.

\section*{Acknowledgements}

I would like to thank S.~Majid for valuable discussions during the
preparation of this paper. I would also like to acknowledge the
financial support by the German Academic Exchange Service (DAAD) and
the Engineering and Physical Sciences Research Council (EPSRC).

\appendix
\section{Definition of $\SUq2$}
\label{app:suq2}

This appendix recalls the defining relations of $\SUq2$ and the
quantum Hopf fibration, see e.g., \cite{Maj:qgroups} or \cite{KlSc:qgroups}.

The matrix Hopf algebra $\SLq2$ is defined over $\C$ with generators
$a,b,c,d$ and
relations
\begin{gather*}
a b = q b a, \quad a c = q c a, \quad b d = q d b,
\quad c d = q d c, \quad b c = c b,\\
a d - d a =(q-q^{-1})b c, \quad a d -q b c = 1,\\
\cop \begin{pmatrix} a & b \\ c & d \end{pmatrix}
= \begin{pmatrix} a & b \\ c & d \end{pmatrix} \dot{\tens}
\begin{pmatrix} a & b \\ c & d \end{pmatrix} \qquad
\cou \begin{pmatrix} a & b \\ c & d \end{pmatrix} =
\begin{pmatrix} 1 & 0 \\ 0 & 1 \end{pmatrix} \\
\antip \begin{pmatrix} a & b \\ c & d \end{pmatrix}
= \begin{pmatrix} d & -q^{-1} b \\ -q c & a \end{pmatrix} .
\end{gather*}
Matrix multiplication is understood in the definition
of the coproduct. The $*$-structure defining the real form $\SUq2$
for real $q$ is given by
\[
 \begin{pmatrix} a & b \\ c & d \end{pmatrix}^*=
 \begin{pmatrix} d & -q c \\ -q^{-1} b & a \end{pmatrix} .
\]

As a Hopf $*$-algebra, $\CZ$ has one generator $g$ with inverse $g^{-1}$
and relations and $*$-structure
\[
 \cop g=g\tens g, \quad \cou g =1,
 \quad \antip g = g^{-1}, \quad g^*=g^{-1} .
\]

There is a Hopf $*$-algebra surjection $\pi:\SUq2\to\CZ$
defined by 
\[
 \begin{pmatrix} a & b \\ c & d \end{pmatrix}\mapsto
 \begin{pmatrix} g & 0 \\ 0 & g^{-1} \end{pmatrix} .
\]
This determines the quantum 2-sphere $\Sq2$ as a right quantum homogeneous
$*$-space under $\SUq2$. At $q=1$ we recover the ordinary Hopf fibration.

\section{Coquasitriangular Structure of $\SUq2$}
\label{app:cqtr}

In this appendix we provide the formulas for the
coquasitriangular structure of $\SUq2$
in the Peter-Weyl basis needed in
Section~\ref{sec:phi4}. 
We use the context of Section~\ref{sec:suq2}.
Definitions and results that are just stated are standard and can be
found e.g.\ in \cite{Maj:qgroups} or \cite{KlSc:qgroups}.

The Hopf algebra $\Uqsl2$ is defined over $\C$ for $q\in\C^*$ and
$q^2\neq 1$ with generators $E,F,K,K^{-1}$ and relations
\begin{gather*}
K E K^{-1}=q^2 E, \quad
K F K^{-1} = q^{-2} F, \\
K K^{-1} = K^{-1} K = 1, \quad 
[E,F]=\frac{K-K^{-1}}{q-q^{-1}},\\
\cop(E)=E\tens K+1\tens E,
\quad \cop(F)=F\tens 1+K^{-1}\tens F,\\
\cop(K)=K\tens K, \quad \cou(K)=1, \quad \cou(E)=\cou(F)=0,\\
\antip(K)=K^{-1}, \quad \antip(E)=-E K^{-1},
\quad \antip(F)=-K F .
\end{gather*}

$\Uqsl2$ and $\SUq2$ are non-degenerately paired.
Thus, actions
of $\Uqsl2$ and coactions of $\SUq2$ on finite dimensional
vector spaces are dual to each other. In particular, the simple
comodule $V_l$ of $\SUq2$ is a simple module of $\Uqsl2$. By the
representation theory of $\Uqsl2$
it has a basis $\{w_i\}$, $i=-l,-l+1,\ldots,l$ such that
\begin{equation}
\begin{gathered}
 K\act w_m=q^{2m} w_m,\quad
 E\act w_m=(\qint{l-m}\qint{l+m+1})^{1/2} w_{m+1}\\
 F\act w_m=(\qint{l+m}\qint{l-m+1})^{1/2} w_{m-1} .
\end{gathered}
\label{eq:usl2act}
\end{equation}

$\Uqsl2$ has an $h$-adic version $\Uhsl2$ defined over $\C[[h]]$
correspondingly with $q=e^h$ and an additional generator $H$ so that
$q^H=K$. It has the quasitriangular structure
\begin{equation}
 R=q^{(H\tens H)/2} \sum_{n=0}^\infty \frac{q^{n(n+1)/2}(1-q^{-2})^n}
 {\qint{n}!} E^n\tens F^n .
\label{eq:sl2qtr}
\end{equation}
The elements (define $R^{(1)}\tens R^{(2)}=R$)
\begin{equation}
 u'=(\antip R^{(2)}) R^{(1)}, \quad v'=R^{(1)}\antip R^{(2)}
\label{eq:defqtruv}
\end{equation}
act on $V_l$ as \cite[Proposition~3.2.7]{Maj:qgroups}
\begin{equation}
 u'\act w_m=q^{-2l(l+1)+2m} w_m,\quad
 v'\act w_m=q^{-2l(l+1)-2m} w_m .
\label{eq:usl2uv}
\end{equation}

The coquasitriangular structure $\Ru$ of $\SUq2$ is given by the
duality with $\Uqsl2$ from the quasitriangular structure $R$ of
$\Uhsl2$.
Using
\[
 u(a\i1) a\i2=\antip^2 a\i1 u(a\i2)\quad\text{and}\quad
 v(a\i1) \antip^2 a\i2=a\i1 v(a\i2)
\]
we find 
\begin{equation}
 u\left(t^{(l)}_{m\,n}\right)
 =\delta_{m,n}\, q^{2(m-k)} u\left(t^{(l)}_{k\,k}\right),\quad
 v\left(t^{(l)}_{m\,n}\right)
 =\delta_{m,n}\, q^{2(k-m)} v\left(t^{(l)}_{k\,k}\right) .
\label{eq:uvid}
\end{equation}
Since the definitions (\ref{eq:defcqtruv}) and (\ref{eq:defqtruv}) are
dual to each other we can use
\[
 g\act v_n=\sum_m v_m \langle g,
 t^{(l)}_{m\,n}\rangle,\quad g\in\Uqsl2
\]
to compare (\ref{eq:usl2uv}) with (\ref{eq:uvid}). We
find (\ref{eq:sl2uv}) and infer that
$w_i$ is (a multiple of) $v_i$.
With the latter, the pairing between $\Uqsl2$ and $\SUq2$ comes out
from (\ref{eq:usl2act}) as
\begin{gather*}
\langle K, t^{(l)}_{m\,n}\rangle=\delta_{m,n}\,q^{2n}, \quad
\langle E, t^{(l)}_{m\,n}\rangle=\delta_{m,n+1}
(\qint{l-n}\qint{l+n+1})^{1/2}, \\
\langle F, t^{(l)}_{m\,n}\rangle=\delta_{m,n-1}
(\qint{l+n}\qint{l-n+1})^{1/2} .
\end{gather*}
Note also $\langle H, t^{(l)}_{m\,n}\rangle=\delta_{m,n} 2n$ in the
$h$-adic version. With this pairing and (\ref{eq:sl2qtr}) we easily
verify the property (\ref{eq:sl2prop}).

\bibliographystyle{amsplainx}
\bibliography{stdrefs}
\end{document}